\documentclass[preprint,12pt]{aastex}
\usepackage{natbib}
\usepackage{amsmath}
\usepackage{epsfig}
\begin{document}

\title{Observational upper limits on the gravitational wave production of core collapse supernovae}

\author{ {Xing-Jiang Zhu\altaffilmark{\dag,\ddag}\footnote{E-mail: zhuxingjiang@gmail.com (ZXJ)}}, {E. Howell\altaffilmark{\dag} and D. Blair\altaffilmark{\dag}} }

\altaffiltext{\dag}{School of Physics, University of Western
Australia, Crawley WA 6009, Australia}
\altaffiltext{\ddag}{Department of Astronomy, Beijing Normal
University, Beijing 100875, China}

\begin{abstract}
The upper limit on the energy density of a stochastic gravitational
wave (GW) background obtained from the two-year science run (S5) of
the Laser Interferometer Gravitational-wave Observatory (LIGO) is
used to constrain the average GW production of core collapse
supernovae (ccSNe). We assume that the ccSNe rate tracks the star
formation history of the universe and show that the stochastic
background energy density depends only weakly on the assumed average
source spectrum. Using the ccSNe rate for $z\leq10$, we scale the
generic source spectrum to obtain an observation-based upper limit
on the average GW emission. We show that the mean energy emitted in
GWs can be constrained within $< (0.49-1.98)\hspace{1mm} M_{\odot}
c^{2}$ depending on the average source spectrum. While these results
are higher than the total available gravitational energy in a core
collapse event, second and third generation GW detectors will enable
tighter constraints to be set on the GW emission from such systems.
\end{abstract}


\section{Introduction}
Core collapse supernovae (ccSNe), neutron star (NS) births and black
hole (BH) births have long been considered to be likely
observational sources of gravitational waves (GWs). For many decades
effort has gone into estimating the GW emission from such sources
(see, e.g., \citealt{thorn,houser,fryer,rezzolla}). Approaches to
the problem have included numerical modelling of ccSNe, study of
normal modes in newly born NSs and BHs, and constraints based on the
observed asymmetry of supernovae or NS kick velocities.

Early estimates for the GW production were $\sim$ a few percent of
rest mass energy, and estimates of gravitational energy loss from
bar mode instabilities in newly born NSs were also quite high.
Recent ccSNe modeling gives much lower predictions of GW production.
For core collapse leading to NS births, predictions for the total
energy emitted in GWs have been in the range $\sim(10^{-12} -
10^{-4})\hspace{1mm} M_{\odot} c^{2}$ \citep{ott}. For BH births the
conversion efficiency to GWs is of the order $\sim(10^{-7} -
10^{-6}) M_{\odot} c^{2}$ \citep{Baiotti}. However, there is
significant uncertainty in supernova modelling due to the incomplete
understanding of the explosion mechanism and the complexity of the
physics involved (see \citet{ott} for a comprehensive review).

It is generally agreed that the frequency of GW emission from the
birth of stellar mass collapsed objects is in the range 50Hz to a
few kHz \citep{muller,DFM,DFM2}. On the other hand, predictions of
GW waveforms from ccSNe depend strongly on the considered emission
mechanism. We argue below that a generic broad Gaussian spectrum
provides a suitable average source spectrum.

Studies of the star formation rate in the universe allow estimation
of the birth rate of collapsed objects. This rate is $\sim 10^2
\hspace{1mm} \rm{s}^{-1}$ as discussed below. This is sufficient to
create a quasi-continuous stochastic gravitational wave background
(SGWB) of astrophysical origin in the above frequency band, with
significant smearing to lower frequencies due to redshift. The
amplitude of this background depends overwhelmingly on the average
GW production for the individual events that combine to create this
background.

The first generation interferometric GW detectors, including LIGO
\citep{LIGO}, Virgo \citep{virgo}, GEO600 \citep{geo} and TAMA300
\citep{tama} have operated for long periods at astrophysically
significant sensitivity \citep{IFOs}. They have enabled constraints
to be placed on various sources of GWs, e.g., pulsars
\citep{crab,pulsar}, deformed NSs \citep{ns}, gamma ray bursts
\citep{SGR,GRB_S5/VSR1} and coalescing compact binaries
\citep{CBC_search}. In particular, blind all-sky burst searches with
LIGO, Virgo and GEO600 detectors set limits for nearby ccSNe that
exploded during data taking \citep{burst_S5y2/VSR1}. In this regard,
LIGO and Virgo are currently developing searches for GW bursts
triggered by gamma ray, optical, radio and neutrino transients.

Recently the LIGO Scientific Collaboration and Virgo Collaboration
used data from a two-year science run (S5) to constrain the energy
density of SGWB in the frequency band $(41.5-169.25)
\hspace{1mm}\rm{Hz}$ to be $<6.9 \times 10^{-6}$ \citep[denoted as
LV limit;][]{sn_limit}. This was interpreted in terms of a
cosmological background due to the big bang and exceeds the previous
indirect limits from the big bang nucleosynthesis and cosmic
microwave background at around 100 Hz.

It has been realized for some time that an astrophysical background
of GWs from multiple individual sources could obscure the
cosmological background (see, e.g.,
\citealt{blair,Ferrari_SNBG_98,coward_GWBG_01a,howell_GWBG_04} among
others). For this reason it is interesting to use the above
observational limit to set a limit on the average GW production from
the most numerous likely sources, ccSNe. To obtain this limit we
consider three different generic average source spectra and work
backwards from the LV limit to estimate an upper limit for the
average GW energy production of ccSNe. First we estimate the cosmic
ccSNe event rate, then discuss the source spectrum, and finally
compare scaled background spectra with the LV limit.

\section{Core collapse supernovae event rate}
The cosmic star formation rate (CSFR) is reasonably well known at
redshifts $z\leq 6$ \citep[][HB06]{SFR}. Since the evolving rate of
ccSNe closely tracks the CSFR, we can estimate the number of ccSNe
events per unit time within the comoving volume out to redshift z:
\begin{equation}
R(z)=\int_{0}^{z}\dot{\rho}_{\ast}(z') \frac{dV}{dz'} dz' \int
\Phi(m)dm \label{dR},
\end{equation}
where $\dot{\rho}_{\ast}(z)$ is the CSFR density in $M_{\odot}
\hspace{0.5mm}\rm{yr}^{-1} \hspace{0.5mm}\rm{Mpc}^{-3}$, $dV/dz$ is
the comoving volume element, and $\Phi(m)$ is the stellar initial
mass function (IMF). For $\dot{\rho}_{\ast}(z)$ and for $dV/dz$ we
use the parametric form given in HB06 and the standard form as in
\citet{Regimbau} respectively. To be consistent with HB06 we also
consider the modified Salpeter A IMF \citep{IMF} assumed to be
universal and independent of z \citep{imf_z}.

We assume that each ccSN results in either a NS or a BH, and
integrate Eq. (\ref{dR}) over a NS (ccSNe) progenitor mass range of
$8 M_{\odot}-25 \hspace{1mm}(100) M_{\odot}$ \citep{mass_range}.
Fig. 1 shows the evolving ccSNe event rate as well as the NS
formation rate. For the total rate of ccSNe and NS formation out to
$z=10$ a maximum value is obtained as 85 $\rm{s}^{-1}$ and 69
$\rm{s}^{-1}$ respectively, with negligible contribution for $z>6$.

\section{Stochastic gravitational wave background}
To obtain the stochastic background, besides knowing the GW event
rate we still need the average source spectrum. The energy flux per
unit frequency emitted by a source at luminosity distance $d_{L}(z)$
is:
\begin{equation}
f(\nu_{\rm{obs}},z)=\frac{1}{4 \pi d_{L}(z)^{2}}
\frac{dE_{\rm{GW}}}{d\nu}(1+z) \label{flux},
\end{equation}
where $dE_{\rm{GW}}/{d\nu}$ is the spectral energy density and $\nu$
is the frequency in the source frame which is related to the
observed frequency by $\nu=\nu_{\rm{obs}}(1+z)$.

Combining Eq. (\ref{dR}) and Eq. (\ref{flux}) we can obtain the
closure energy density $\Omega_{\rm{GW}}$ of the SGWB:
\begin{equation}
\Omega_{\rm{GW}}(\nu_{\rm{obs}})=\frac{\nu_{\rm{obs}}}{c^{3}\rho_{c}}\int_{0}^{10}
[f(\nu_{\rm{obs}},z)(dR/dz)]dz,
\end{equation}
where $\rho_{c}$ is the cosmological critical energy density.

One particular obstacle in estimating the SGWB comes from
uncertainties in the source spectra. Fortunately, for a stochastic
background detailed structures of source spectra are not important
due to the following two facts.

Firstly the smearing of the spectrum by redshift means that a SGWB
cannot be sharply peaked. Any narrow spectral details are greatly
broadened. For example, if all sources emitted a narrow spectrum at
1 kHz, the resulting stochastic background is broadly peaked at 600
Hz with half-width $\sim600$ Hz assuming our source rate evolution
model. When individual sources create a stochastic background,
redshift washes out most of the detailed structure.

Secondly a stochastic background consists of an average over a very
large number of sources. For example, the frequency spectrum of a
SGWB obtained from one-year cross correlation would be produced by
the superposition of more than $10^9$ individual events. Given that
source spectra are likely to vary with progenitor masses and angular
momentum, any fine spectral detail will be averaged. In addition the
BH formation spectrum scales inversely with BH mass, so again the
average source will be smoothed and broadened assuming a continuous
distribution of BH masses.

For the above two reasons a much simpler generic spectrum can be
adopted for an average source. We go on to show that a Gaussian
spectrum is a suitable approximation where $dE_{\rm{GW}}/{d\nu}$ is
given by:
\begin{equation}
{dE_{\rm{GW}}\over{d\nu}}= A \exp
\bigg[-\frac{(\nu-\nu_0)^2}{2{\Delta}^2}\bigg] \label{gaussian},
\end{equation}
with amplitude $A$, peak frequency $\nu_0$, half-width $\Delta$.

Many of the modelling predictions show a range of spectral density
varying over many orders of magnitude. However only the largest
peaks in the source spectra contribute significantly to the
stochastic background. For most predicted source waveforms one can
obtain a very similar stochastic background from a Gaussian source
spectrum. We have compared the resultant SGWB from 72 source models
\footnote{Data are taken from http://www.stellarcollapse.org/} of
\citet{ott04} with the signals obtained using a Gaussian
approximation. We find, through comparison of the frequency
integrated energy in the peak decade, that $87\%$ of our models are
consistent to within $10\%$.

For illustration, in Fig. 2. we determine the SGWB based on two
different GW emission mechanisms:  firstly rotating core-collapse
and bounce based in model s11 of \citet{ott04}; secondly post
collapse emission from accretion and turbulence driven proto-NS
oscillations (acoustic mechanism) using model s11WW of
\citet{ott06}. The plot shows that for the highest decade, a
Gaussian source spectrum given by Eq. (\ref{gaussian}) with
$\nu_0=175$ Hz, $\Delta=120$ Hz (model 1) can produce a GW
background with a spectral distribution comparable to that obtained
using s11 (with a relative error $1.9\%$). Similarly, the SGWB
determined using s11WW can also be well estimated (within $16.1\%$)
by using a source model 2, with $\nu_0=860 \hspace{1mm}\rm{Hz},
\Delta=130\hspace{1mm}\rm{Hz}$. Although the exact spectral shape is
not reproduced for model 2, we argue that the precise shape of the
SGWB is not essential to obtain upper limits.

For above reasons in this paper, rather than adopt predicted spectra
and determine the GW background, we have chosen to adopt a generic
spectrum which takes into account the range of spectral predictions
and use this to obtain an observational upper limit on the average
GW production of ccSNe. Taking into account the range of progenitor
masses and angular momentum, we would expect that the average source
spectrum should be wider than those corresponding to particular
simulations.

Based on the simulated ccSNe spectra of \citet{DFM2} (see Table 1 of
\citet{ott}); and on spectra resulting from BH births from
\citet{sekiguchi}, for which typical peak frequencies occur at
around 1 kHz; we consider three Gaussian source spectra with the
following parameters: a) $\nu_0=200 \hspace{1mm}\rm{Hz}, \Delta=200
\hspace{1mm}\rm{Hz}$; b) $\nu_0=500 \hspace{1mm}\rm{Hz},
\Delta=400\hspace{1mm}\rm{Hz}$ and c) $\nu_0=800
\hspace{1mm}\rm{Hz}, \Delta=500 \hspace{1mm}\rm{Hz}$ (denoted as
model a, b and c respectively). In the next section we will use the
above three generic source models to compare the SGWB from
cosmological ccSNe with the LV limit.

\section{Upper limits on the gravitational wave production of core collapse supernovae}
We define the average GW production of ccSNe as:
$E_{\rm{GW}}=\epsilon \cdot M_{\odot} c^2$ where $\epsilon$ is the
average energy production in solar rest mass units and the total
energy $E_{\rm{GW}}$ is distributed as Eq. (\ref{gaussian}). We
obtain an upper limit on $\epsilon$ by scaling the SGWB calculated
from a Gaussian average source model to produce a comparable
signal-to-noise ratio (SNR) to that of a frequency-independent flat
GW background. Assuming Gaussian noise in each of two
cross-correlated detectors separated by less than one reduced
wavelength, the optimal SNR after correlating outputs of two
detectors during an integration time T is given by an integral over
frequency $f$ \citep[][Eq. 3.75]{SNR}:

\begin{equation}
\left( {S \over N} \right)^2 = {9 H_0^4 \over 50 \pi^4} T
\int_0^\infty df \> {\gamma^2 (f) \Omega_{\rm{GW}}^2(f) \over f^6
P_1(f) P_2(f)}\ \label{SNR},
\end{equation}

\noindent with $P_1(f)$ and $P_2(f)$ the power spectral noise
densities of the two detectors and $\gamma(f)$ the overlap reduction
function determined by the relative locations and orientations of
two detectors \citep{gammaf}. Here we adopt an integration time of
292 days and $\gamma(f)$ for LIGO H1-L1 calculated using Eq. (3.26)
of \citet{allen2002}. We calculate Eq. (\ref{SNR}) in the frequency
band $(41.5-169.25) \hspace{1mm}\rm{Hz}$ using representative noise
spectra\footnote{Over long observational periods the detector noise
is non-stationary but can be approximated by the representative
noise spectra of S5 which can be found at
https://dcc.ligo.org/cgi-bin/private/DocDB/ShowDocument?docid=6314}.
We find that the SNR given by Eq. (\ref{SNR}) is 2.58 for a flat
stochastic background $\Omega_{\rm{GW}} = 6.9 \times 10^{-6}$. In
order to achieve the same SNR the average GW production $\epsilon$
is required to be $0.49$, $1.08$ and $1.98$ assuming source model a,
b and c respectively.

Fig. 3 shows the LV limit along with the predicted stochastic
backgrounds from ccSNe for the three average source models. The
strongest constraint is obtained when individual sources emits
gravitational radiation at a peak frequency near LIGO's most
sensitive frequency band. This corresponds to $\epsilon \leq 0.49$
($E_{\rm{GW}}=8.8 \times 10^{53} \hspace{1mm} \rm{ergs}$) assuming
an average source spectrum given by model a. Note that this limit is
higher than the available energy ($\sim 3.0 \times 10^{53}
\hspace{1mm} \rm{ergs}$) for any kind of emission in a ccSNe event
which is set by the binding energy of the final NS. If the source
peak frequency increases, as in models b and c we obtain somewhat
higher limits (1.08 and 1.98 respectively) for $\epsilon$. Fig. 3
also illustrates that stronger constraints can be obtained if
stochastic background limits at higher frequencies are available.

It is important to consider the effect of other stochastic
background sources. We can be reasonably certain that the true
stochastic background contains contributions from compact binary
coalescences and individual spinning NSs. All these sources
contribute GW energy in the relevant frequency band. Of these, the
coalescing compact binaries are worth considering because
significant fraction of system binding energy is expected to be
converted to GWs. However, their event rate is much lower, about
$10^{-3}$ of ccSNe rate \citep{sado,lsc_rate,metal}. Thus applying
similar methods to such systems will yield $10^{3}$ higher upper
limit on the average GW production. We therefore only consider ccSNe
in this study.

It is interesting to note that the LV limit is comparable to the
minimum detectable GW energy density with two initial LIGO detectors
predicted some years ago - $\Omega_{\rm{GW}} \sim 5\times 10^{-6}$
\citep[see, e.g.,][]{allen,Maggiore}. A world-wide network of
advanced detectors have been planned or proposed, including Advanced
LIGO \citep{aligo}, Advanced Virgo \citep{avirgo}, LCGT \citep{lcgt}
and AIGO \citep{aigo}. Additionally design studies for a
third-generation GW observatory, Einstein Telescope
\citep[ET;][]{et}, are well underway with a target sensitivity 100
times better than current instruments. While limits obtained here
are far away from being able to test GW emission mechanisms of
ccSNe, future measurements will lead to great improvements in this
regard. For instance, Advanced LIGO will be able to detect or to set
a much stronger limit on the stochastic background at the level of
$\Omega_{\rm{GW}} \geq 9 \times 10^{-10}$ \citep{advanced}. This
will imply an upper limit on the GW production of ccSNe as low as
$\sim 10^{-5} \hspace{1mm} M_{\odot} c^2$. For ET, the minimum
detectable value $\Omega_{\rm{GW}} \sim 5 \times 10^{-12}$
\citep{SathyaprakashSchutz_LIVREV} corresponds to a limit $\sim
10^{-7} \hspace{1mm} M_{\odot} c^2$, which is in the range of
predictions from various ccSNe simulations.

\section{Conclusions}
We have shown that upper limits  on the stochastic background of GWs
can be used to set limits on the GW energy production in ccSNe. This
result is the first upper limit for GW production averaged over all
core collapse events out to z $\sim10$. The upper limit is in the
range $(0.49-1.98) \hspace{1mm} M_{\odot} c^2$ depending on the
average source spectrum. Our result represents an average over both
space and cosmic time, thereby including an average over possible
evolutionary effects in GW production. It is higher than the upper
limit on the available energy for explosion in a core collapse
event. However second and third generation GW detectors will enable
tighter constraints to be set on the GW emission from such systems.
Using our methods the predicted upper limits on the average GW
production from ccSNe will be $10^{-5} \hspace{1mm} M_{\odot} c^2$
and $10^{-7} \hspace{1mm} M_{\odot} c^2$ for Advanced LIGO and ET
respectively.

Compact binary coalescence events are not considered in the above
limit because their event rate is very low in comparison with ccSNe.
However, at the expected improved limits from Advanced LIGO and ET,
it will be possible to consider GW backgrounds from other sources
like compact binary coalescences.

It would be interesting to compare our upper limits with those on
the fraction of stellar core rest mass converted to GWs obtained by
averaging GW detector outputs over time periods associated with
nearby ccSNe \citep{finn}. Such methods have already been used to
set upper limits on the total energy emitted in GWs for one
particular astrophysical event during searches for GW bursts. For
example, \citet{GRB} placed an upper limit of $1.6 \times 10^4
\hspace{1mm} M_{\odot} c^2$ on the total GW energy for GRB 050223 (D
$\approx 3.5 \hspace{1mm} \rm{Gpc}$) which was considered to be
associated with a core collapse. In \citet{GRB_S5/VSR1}, a GW energy
limit for GRB 070201 was given as $1.15 \times 10^{-4} \hspace{1mm}
M_{\odot}$ at 150 Hz assuming a position of M31 (770 kpc). As we can
see such limits depend strongly on the knowledge of source distance.
\citet{burst_S5y2/VSR1} set a limit on the total energy of one GW
burst event (assuming a sine-Gaussian signal) that would be
detectable with the current LIGO-Virgo detectors as $1.8 \times
10^{-8} \hspace{1mm} M_{\odot} c^2$ and $4.6 \times 10^{-3}
\hspace{1mm} M_{\odot} c^2$ for a typical Galactic distance (10 kpc)
and the Virgo cluster (16 Mpc) respectively.

This method suffers from the poor time resolution for the collapse
event (unless timed by detected neutrinos\footnote{There is a
significant ongoing effort to forge a collaboration between
LIGO-Virgo detectors and neutrino detectors in order to carry out a
neutrino-triggered search for GWs from ccSNe
\citep{h_neutrino,l_neutrino}.}) and a small sample of events during
the period that GW data are available. However with the improvements
of detector sensitivities, stronger upper limits or even positive
detection of GWs from ccSNe will be possible based on single or
small numbers of events in the future.

\section*{Acknowledgments}
We acknowledge the US National Science Foundation, the LIGO
Scientific Collaboration, and the Virgo Collaboration for providing
S5 noise spectra data. The authors are grateful to Xi-Long Fan, Dave
Coward and Linqing Wen for helpful discussions, and to Prof.
Zong-Hong Zhu for his support of ZXJ's visit at UWA. The authors
also gratefully acknowledge Christian Ott and Peter Kalmus for
insightful comments which have led to some valuable amendments. Our
thanks also go to the anonymous referee for useful comments which
have improved the clarity of this letter. This research was funded
by the Australian Research Council and the W.A. Government Center of
Excellence Program. ZXJ is supported in part by the National Science
Foundation of China under the Distinguished Young Scholar Grant
10825313 and by the Ministry of Science and Technology national
basic science Program (Project 973) under grant No. 2007CB815401.

\begin{figure}
\plotone{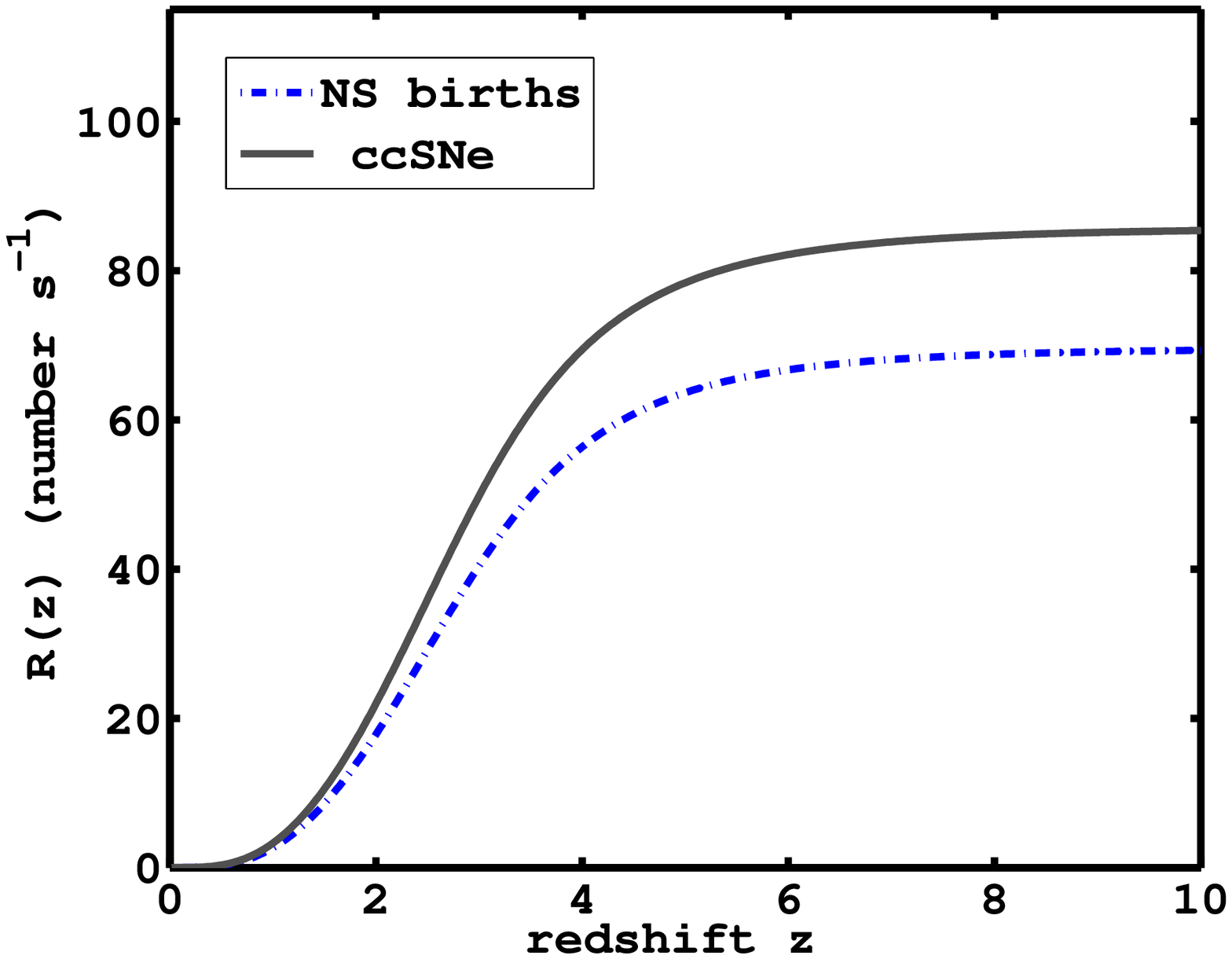} \figcaption{The total number of core
collapse supernovae (ccSNe) events and neutron star (NS) births
occurring per unit time within the comoving volume out to redshift
z.}
\end{figure}

\begin{figure}
\plotone{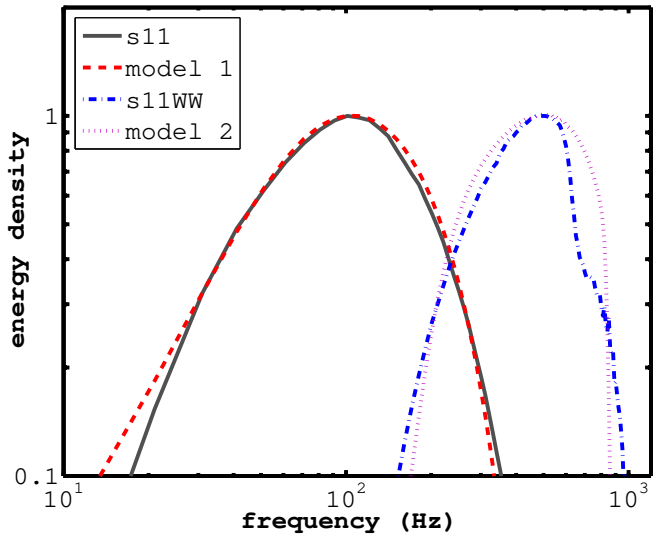} \figcaption{The relative spectral
distribution of the SGWB from NS core collapse assuming two Gaussian
source spectra (model 1 and model 2 ) and two simulated waveforms:
s11A500B0.5 (denoted as s11) from \citet{ott04} and s11WW from
\citet{ott06}. Here s11 indicates a $11 M_{\odot}$ presupernova
model and we refer readers to the original papers for details of the
simulations. We adopt the NS formation rate in Fig. 1 and scale the
energy density of the SGWB to an arbitrary logarithmic scale. As
only the strongest part of the background is essential to estimate
an upper limit, we show only the highest decade of the relative
stochastic background.}
\end{figure}

\begin{figure}
\plotone{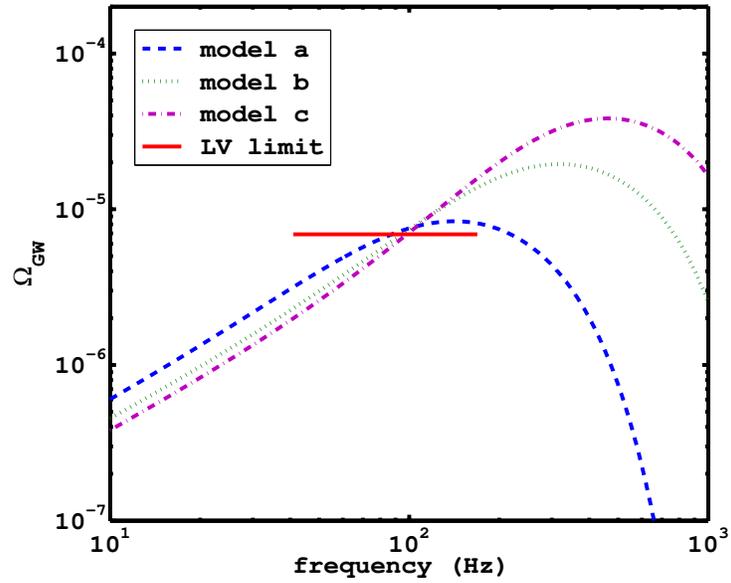} \figcaption{The upper limit on SGWB and
GW background from cosmological ccSNe assuming three different
Gaussian source models (see text). The models have been scaled to
produce the same SNR within (41--169) Hz as that of a flat
frequency-independent background (LV limit).}
\end{figure}

\end{document}